\documentclass{jkas}

\def\year{2021} 
\def\month{June} 
\def\volume{54} 
\def\issue{3} 
\def\beginpage{103} 
\def\endpage{112} 
\def\received{May 4, 2021} 
\def\accepted{June 15, 2021} 
\date{Received \received; accepted \accepted}

\def\eg{{e.g.,\ }}
\def\kms{~{\rm km~s^{-1}}}
\def\cm3{~{\rm cm^{-3}}}
\def\yrs{~{\rm yrs}}
\def\muG{~{\mu\rm G}}

\usepackage{flushend}

\title{Diffusive Shock Acceleration by Multiple Weak Shocks}

\author{Hyesung Kang}

\affil{Department of Earth Sciences, Pusan National University, Busan 46241, Korea; \email{hskang@pusan.ac.kr}}

\def\corrauthor{H. Kang}
\def\runningauthor{Kang}
\def\runningtitle{DSA by Multiple Weak Shocks}
\def\keywords{acceleration of particles --- cosmic rays ---
galaxies: clusters: general --- shock waves}

\def\abstracttext{
The intracluster medium (ICM) is expected to experience on average about three passages of weak shocks with
low sonic Mach numbers, $M\lesssim 3$, during the formation of galaxy clusters.
Both protons and electrons could be accelerated to become high energy cosmic rays (CRs)
at such ICM shocks via diffusive shock acceleration (DSA).
We examine the effects of DSA by multiple shocks on the spectrum of accelerated CRs by
including {\it in situ} injection/acceleration at each shock, followed by
repeated re-acceleration at successive shocks in the test-particle regime.
For simplicity, the accelerated particles are assumed to undergo adiabatic decompression without energy loss
and escape from the system, before they encounter subsequent shocks.
We show that in general the CR spectrum is flattened by multiple shock passages, compared to a single episode of DSA,
and that the acceleration efficiency increases with successive shock passages.
However, the decompression due to the expansion of shocks into the cluster outskirts may 
reduce the amplification and flattening of the CR spectrum by multiple shock passages.
The final CR spectrum behind the last shock is determined by the accumulated effects of repeated 
re-acceleration by all previous shocks, but it is relatively insensitive to the ordering of the shock Mach numbers.
Thus multiple passages of shocks may cause the slope of the CR spectrum to deviate from
the canonical DSA power-law slope of the current shock.}
\begin{document}
\jkashead 
\section{Introduction}

Cosmic rays (CRs) are known to be produced primarily
via diffusive shock acceleration (DSA) in a variety of astrophysical shocks  \citep{bell78,blaeic87}.
In the test particle regime, where the CR pressure is dynamically insignificant,
DSA predicts the power-law momentum distribution 
of accelerated particles, $f(p) \propto p^{-q}$,
where the slope is $q=3r/(r-1)$ and $r$ is the shock compression ratio \citep{drury1983}.  
For strong, adiabatic, non-relativistic shocks, this would give a power-law index of $q=4$,
if nonlinear DSA effects were to be ignored.
However, plasma simulations of collisionless shocks demonstrated that at strong quasi-parallel shocks 
order of 10 $\%$ of the shock kinetic energy could be transferred 
to CRs \citep[e.g.,][]{caprioli14,park2015}, 
possibly leading to nonlinear back-reactions from CRs to the underlying flow.

Several previous studies 
considered multiple passages of shocks as a possible scenario to achieve a CR spectrum
flatter than $p^{-4}$ \citep[e.g.,][]{white1985,achterberg90,schneider1993,melrose1993, gieseler2000}. 
These authors showed that for a large number of {\it identical} shocks, the resulting spectrum
approach to $f(p)\propto p^{-3}$, independent of the shock compression ratio (or shock Mach number).
In a more realistic situation, the injection process at each shock, adiabatic decompression in the far-downstream region, and further accelerations due to magnetic turbulence in the postshock region 
should be considered in addition to
the re-acceleration of incident upstream CRs.
For instance, \citet[][MP93 hereafter]{melrose1993} assumed that mono-energetic particles are 
injected with the spectrum, $\phi_0(p)=K\delta(p-p_0)$, at each shock,
and that the postshock flow is decompressed by a factor of $1/r$ between each shock.
They provided an analytic expression for the distribution function $f(p)$ after multiple passages of identical shocks.

In this paper, we revisit the problem of DSA by multiple shocks to explore the effects of re-acceleration
of CRs by weak shocks that are induced in the intracluster medium (ICM) of galaxy clusters.
Cosmological hydrodynamic simulations demonstrated that the ICM could 
encounter shocks several times on average during the formation 
of the large scale structures in the Universe \citep[e.g.,][]{ryu2003,vazza09, vazza11}.
Such ICM shocks are expected to
produce CR protons and electrons via DSA \citep[see][for a review]{brunetti2014}.
In particular, merger-driven shocks have been detected as radio relics
in the outskirts of galaxy clusters through radio synchrotron radiation from shock-accelerated CR electrons \citep[e.g.][]{vanweeren2019}.
In a recent study, using cosmological hydrodynamic simulation, \citet{ha2020} have estimated that 
the slope of the CR proton spectrum flattens by $\sim0.05-0.1$
and the total energy of CR protons increases by $\sim40-80$\% due to the re-acceleration
of three passages of weak shocks through the ICM throughout the structure formation history.

The physics of collisionless shocks depends on various shock parameters including the sonic Mach number, $M$, the plasma beta, $\beta\equiv P_{\rm g}/P_{\rm B}$, and the obliquity angle, $\theta_{\rm Bn}$, between the upstream background magnetic field direction and the shock normal \citep[e.g.,][]{marcowith2016}.
CR protons are known to be accelerated efficiently at quasi-parallel ($Q_\parallel$) shocks with $\theta_{\rm Bn}\lesssim 45^{\circ}$ \citep[e.g.][]{caprioli15}, 
while CR electrons are accelerated preferentially 
at quasi-perpendicular ($Q_\perp$) shocks with $\theta_{\rm Bn}\gtrsim 45^{\circ}$ \citep[e.g.][]{guo14}.

Recent studies have suggested that in the weakly magnetized ICM plasmas ($\beta~\sim 100$)
the proton acceleration is effective only at supercritical $Q_\parallel$-shocks with 
$M \gtrsim 2.3$ due to the efficient reflection of protons at the shock \citep{ha2018b}.
By contrast, the electron acceleration is effective only at supercritical $Q_{\perp}$-shocks with $M \gtrsim 2.3$ via the excitation of the electron firehose instability \citep{kang2019}
and the Alfv\'{e}n ion cyclotron instability \citep{ha2021}.
Based on these plasma simulation studies, \citet{ryu2019} and \citet{kang2020} proposed analytic forms for 
$f(p)$ for CR protons and electrons, respectively, which are valid for weak shocks in the test-particle regime.
However, it is known that magnetic fluctuations on the relevant kinetic scales 
could facilitate particle injection to DSA \citep[e.g.][]{guo2015}.
So the effects of pre-existing magnetic turbulence in the ICM on DSA at subcritical shocks
($M< 2.3$) have yet to be understood.

As discussed in \citet{kang2020}, DSA can operate in the two different modes:
(1) {\it in situ} injection/acceleration mode, in which particles are injected directly from the background thermal pool at the shock,  and 
(2) re-acceleration mode, in which pre-existing CR particles are accelerated repeatedly at subsequent shocks.
In the test-particle regime, the particle spectrum resulted from these two modes of DSA 
can be found either analytically or semi-analytically.

We defer the exploration of the electron acceleration by multiple shocks to a future work,
since energy losses of CR electrons due to synchrotron radiation and inverse Compton (iC) scattering
introduce additional physical scales. 
However, here we briefly describe a discrepancy related to observations of radio relics in order to allude a possible implication of electron DSA by multiple shocks.
For some `radio relic shocks', the radio Mach number, $M_{\rm rad}=[(\alpha_{\rm int}+1))/(\alpha_{\rm int}-1)]^{1/2}$, 
is higher than the X-ray Mach number, $M_{\rm X}$, inferred from X-ray observations
(e.g., temperature jump),
where $\alpha_{\rm int}=(q-2)/2$ is the spectral index of the integrated radio spectrum \citep[e.g.,][]{akamatsu13, vanweeren2019}.
Possible solutions to explain this puzzle suggested so far include
re-acceleration of preexisting fossil CR electrons with a flat spectrum \citep[e.g.,][]{kang2016} and {\it in situ} acceleration by an ensemble of shocks formed in
the turbulent ICM \citep[e.g.,][]{hong2015,roh2019,rajpu2020, DF2021}.

In the next section we describe the semi-analytic approach to follow DSA by multiple shocks
along with the underlying assumptions and justification.
In Section \ref{s3}, we apply our approach to a few examples, where the re-acceleration by
several weak shocks of $M\le 3$ in the ICM environment is considered.
A brief summary will be given in Section \ref{s4}.

\section{DSA Spectrum by Multiple Shocks}
\label{s1}

The ICM is expected to experience on average about 3 passages of shocks with a mean separation,
$L \sim 1$~Mpc, during the formation of galaxy clusters \citep[e.g.][]{ryu2003}.
The average Mach number of shocks formed in the outskirts of typical galaxy cluster 
is estimated to be in the range $\langle M \rangle \sim 2-2.5$ \citep[e.g.][]{vazza11, ha2018a}.
The mean time between consecutive shock passages with $V_{\rm s} \sim 3\times 10^3 \kms$ 
can be approximated roughly as
$t_{\rm pass} \sim L/V_{\rm s} \sim 3\times 10^8 \yrs$. 

During the time between consecutive shock passages, $t_{\rm pass}$, CR protons can be accelerated via DSA up to the maximum momentum,
\begin{equation}
 {p_{\rm max} \over {m_pc}} \approx  1.5\times 10^{9}
\left({V_{\rm s} \over {3\times 10^3 \kms}}\right)^2 \left({t_{\rm pass} \over {10^8 {\yrs}}}\right)
\left({B_{\rm ICM} \over 1\muG}\right),
\label{pmax}
\end{equation}
where the shock compression ratio, $r=3$, is adopted, and $B_{\rm ICM}$ is the magnetic field strength in the ICM in units of microgauss \citep{kang2020}.
Energy loss processes such as Coulomb collisions, synchrotron emission, and the interactions with background radiation are not considered in the estimation of $p_{\rm max}$ here. 
Throughout the paper, common symbols in physics are used: \eg 
$m_{\rm p}$ for the proton mass, $c$ for the speed of light, 
and $k_{\rm B}$ for the Boltzmann constant.
The gyroradius of CR protons is given as
\begin{equation}
r_{\rm g} (p) \approx 1.1~{\rm kpc}  
\left( p_{\rm max} \over {10^{9}~{m_pc}}\right) \left({B_{\rm ICM} \over 1\muG}\right)^{-1},
\label{rg}
\end{equation}
so typically $r_{\rm g} (p)\ll L$ a the CR proton energy of $E< 10^{18}$~eV.

\subsection{Preambles}
\label{s2.1}

\begin{figure}[t]
\vskip 0 cm
\centerline{\includegraphics[width=0.50\textwidth]{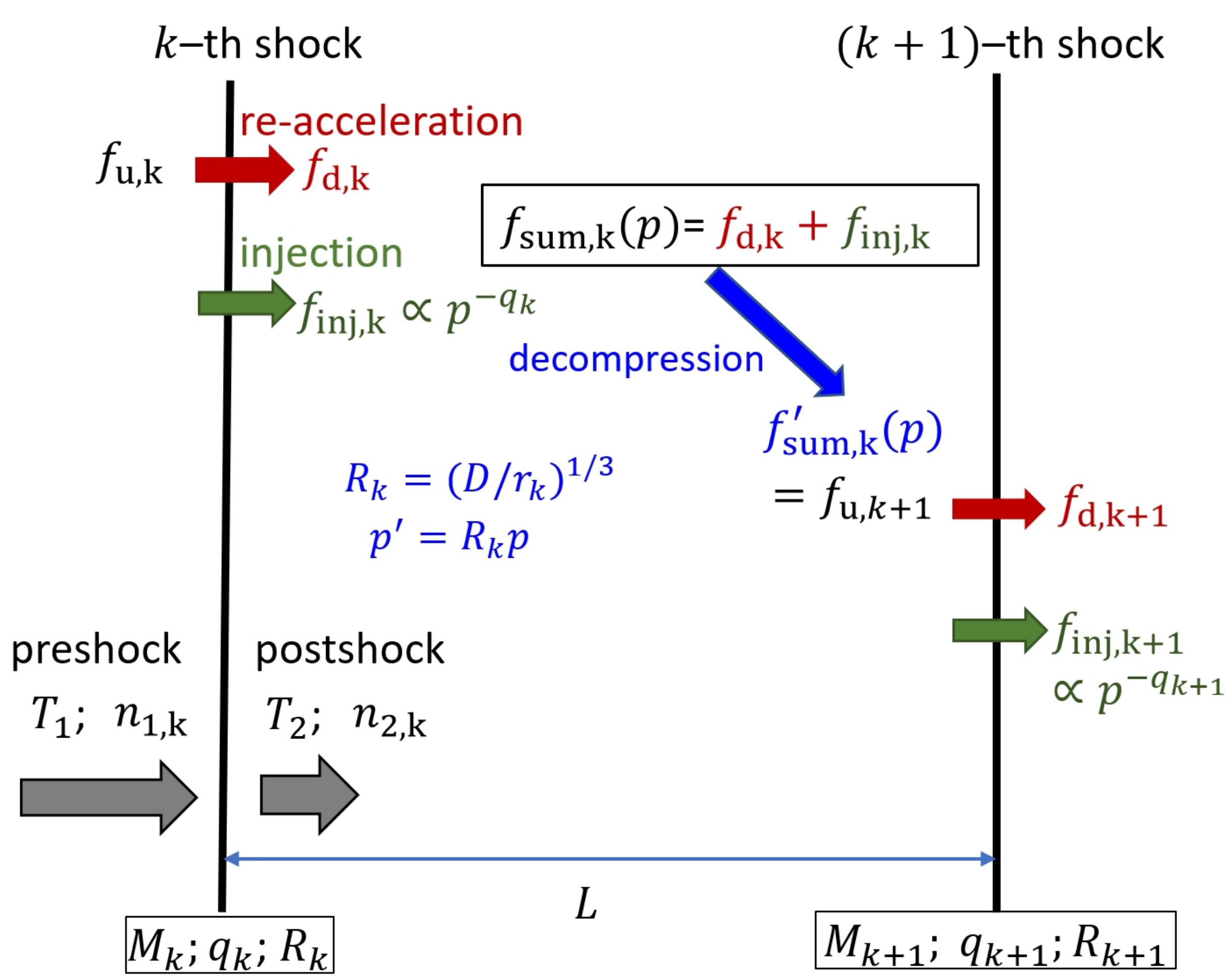}}
\vskip 0 cm
\caption{Basic concept of DSA by multiple shocks adopted in this study. 
The upstream CR spectrum, $f_{\rm u,k}$, is re-accelerated (red arrow) at the $k$-th shock, resulting
in the downstream spectrum, $f_{\rm d,k}$. 
In addition, particles are injected and accelerated (green arrow) to form 
$f_{\rm inj,k}$ at each shock.
Then the downstream spectrum is the sum of the two components, $f_{\rm sum,k}=f_{\rm d,k}+f_{\rm inj,k}$.
After decompression (blue arrow), the particle momentum decrease to $p^{\prime}= R_{\rm k}p$, and so
the decompressed spectrum becomes $f_{\rm sum,k}^{\prime}(p)=f_{\rm sum,k}(p/R_{\rm k})$, where 
$R_{\rm k}=(\mathcal{D}/r_{\rm k})^{1/3}$ is the decompression factor.
The upstream spectrum at the subsequent shock is taken to be $f_{\rm u,(k+1)}= 
f_{\rm sum,k}^{\prime}$.
\label{f1}
}
\end{figure}

We consider a sequence of consecutive shocks that propagate
into the upstream gas of the temperature, $T_1$, and the hydrogen number density, $n_1$.
Hereafter, the subscripts, $1$ and $2$, denote the preshock and postshock states, respectively.
We adopt the following assumptions (see Figure 1):
\begin{enumerate}
\item There are $N_s$ consecutive shocks that sweep the background medium.
The $k$-th shock is specified by a sonic Mach number, $M_{\rm k}$, which determines
the shock compression ratio, $r_{\rm k}=n_{\rm 2,k}/n_{\rm 1,k}$, the slope of DSA power-law spectrum,
$q_{\rm k}= 3 r_{\rm k}/(r_{\rm k}-1)$, and the temperature jump, $T_{\rm 2,k}/T_{\rm 1,k}$.
For simplicity, we presume that the hydrogen number density of the underlying ICM may decrease by a factor of 
$\mathcal{D} \le 1$, that is, $n_{\rm 1,(k+1)}=\mathcal{D} n_{\rm 1,k}$,
considering that the ICM shocks may expand into the stratified outskirts of the host cluster.
In order to limit the number of free parameters, we fix the preshock temperature at $T_1$.

\item Adopting the DSA power-law in the test-particle regime for weak shocks with $M_{\rm k}\lesssim 3$, 
the postshock spectrum of CR protons, injected and accelerated
at the $k$-th shock, is assumed to have the form of $f_{\rm inj,k} \propto p^{-q_{\rm k}}$ 
for $p\ge p_{\rm inj,k}$, where $p_{\rm inj,k}$ is the injection momentum.

\item Even subcritical shocks with $M\lesssim 2.3$ could accelerate CRs via DSA, 
presuming that the ICM contains pre-existing magnetic turbulence on the relevant kinetic scales.
Scattering of particles off such turbulent waves is a prerequisite for DSA.

\item The upstream spectrum of incident particles, $f_{\rm u,k}(p)= f_{\rm sum,k-1}^\prime(p)$, 
includes the contributions of CRs injected and re-accelerated by all previous shocks.
It becomes the downstream spectrum, $f_{\rm d,k}(p)$, after the re-acceleration at the $k$-th shock.  

\item In the postshock region, CRs are transported and decompressed adiabatically without
energy losses (e.g., radiation) and escape from the system,
so the particle momentum, $p$, decreases to $p^{\prime}= R_{\rm k} p$,
where $R_{\rm k}=(\mathcal{D}/r_{\rm k})^{1/3}$ is the decompression factor of the $k$-th shock (see MP93).
Due to this combined decompression, the far-downstream spectrum becomes, $f^{\prime}(p)= f(p/R_{\rm k})$.
As illustrated in Figure \ref{f1}, the total downstream spectrum $f_{\rm sum,k}(p)$
is decompressed to become $f_{\rm sum,k}^\prime(p)$ behind the $k$-th shock.

\item The time between the consecutive passages of the $k$-th and $(k+1)$-th shocks
is much longer than the DSA time scale, so $p_{\rm max,k} \gg p_{\rm inj,k}$.
In other words, the timescale of shock passage can be separated from the DSA timescale.
\end{enumerate}

\subsection{Injected Spectrum at Each Shock}
\label{s2.2}

Following \citet{ryu2019}, we assume that
suprathermal protons with $p \gtrsim p_{\rm inj,k}$ could 
diffuse across the shock transition layer, and participate in the DSA process at the $k$-th shock
with the sonic Mach number $M_{\rm k}$.
Here the injection momentum is defined as
\begin{equation}
p_{\rm inj,k} = Q \cdot p_{\rm th,k},
\end{equation}
where the postshock thermal momentum, $p_{\rm th,k}=(2m_p k_B T_{\rm 2,k})^{1/2}$, depends on the postshock temperature,
$T_{\rm 2,k}$, and the injection parameter is set to be $Q\approx 3.8$
\citep[e.g.,][]{caprioli14, caprioli15,ha2018b}.

At each shock the test-particle spectrum of CR protons results from particle injection to DSA for $p>p_{\rm inj,k}$ with the following power-law form:
\begin{equation}
f_{\rm inj,k}(p) = f_{\rm o,k} \cdot \left(p \over p_{\rm inj,k} \right) ^{-q_{\rm k}}.
\label{finj}
\end{equation}
In this study, thermal protons are assumed to follow a Maxwellian distribution for $p\le p_{\rm inj,k}$,
so the normalization factor is specified at $p_{\rm inj,k}$ as
\begin{equation} 
f_{\rm o,k} = {n_{\rm 2,k} \over \pi^{1.5}} p_{\rm th,k}^{-3} \exp(-Q^2),
\label{fo}
\end{equation}
where $n_{\rm 2,k}$ is the postshock proton number density.
The far-downstream spectrum of injected/accelerated CRs, decompressed behind the $k$-th shock, becomes
\begin{equation}
f_{\rm inj,k}^{\prime}(p) = f_{\rm o,k} \cdot \left(p/R_{\rm k} \over p_{\rm inj,k} \right) ^{-q_{\rm k}}.
\label{fpinj}
\end{equation}
Hereafter, the primed distribution function, $f^{\prime}(p)$, represents the decompressed spectrum 
in the far-downstream region of each shock.

If the acceleration is limited by a finite size or age, 
an exponential cutoff at a maximum momentum, $p_{\rm max}$,
should be applied to Equations (\ref{finj}) and (\ref{fpinj}).
We do not consider such cases, since $p_{\rm max}\gg m_pc$.


\begin{figure*}[t]
\vskip -0.0 cm
\centerline{\includegraphics[width=0.70\textwidth]{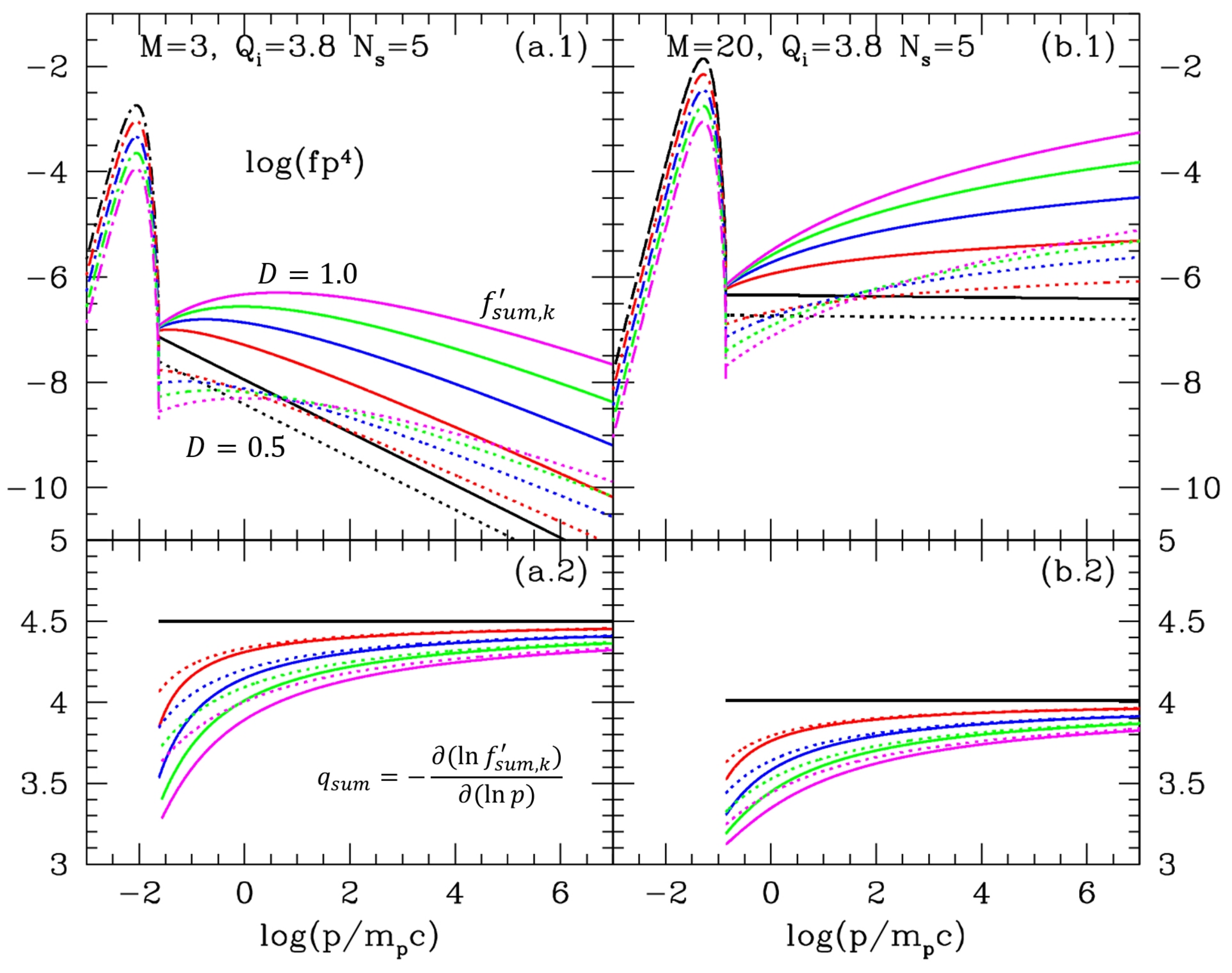}}
\vskip -0 cm
\caption{DSA at five identical shocks with $M=3$ (left) and $M=20$ (right).
Panels (a.1) and (b.1) show $f_{\rm sum,k}^{\prime}(p)$ with $\mathcal{D}=1$ (solid lines) and $\mathcal{D}=0.5$ (dotted lines). The dashed lines show the Maxwellian distributions for the cases with $\mathcal{D}=0.5$.
Panels (a.2) and (b.2) show the power-law slope,
$q_{\rm sum}= - \partial ( \ln f_{\rm sum,k}^{\prime})/\partial \ln p$.
The injection parameter is set to be $Q_{\rm i}=3.8$.
The black, red, blue, green, and magenta lines are used for $k=$1, 2, 3, 4, and 5, respectively.
The cases with $M=20$ are shown for illustrative purposes only,
since the test-particle assumption is not valid for such strong shocks. 
\label{f2}
}
\end{figure*}

\subsection{Re-acceleration by Subsequent Shocks}
\label{s2.2}

The upstream spectrum of the $k$-th shock, $f_{\rm u,k}(p)$, contains
the particles injected and re-accelerated by all previous shocks, which are convected downstream and decompressed.
Then, the downstream spectrum, $f_{\rm d,k}$, re-accelerated at the $k$-th shock, 
can be calculated by the following integration \citep{drury1983}:
\begin{equation}
f_{\rm d,k}(p,p_{\rm inj,k})= q_{\rm k} \cdot p^{-q_{\rm k}} \int_{p_{\rm inj,k}}^p t^{q_{\rm k}-1} f_{\rm u,k} (t) dt,
\label{reacc}
\end{equation}
where the lower bound is set to be $p_{\rm inj,k}$, above which the particles can 
participate in DSA.
Then the immediate postshock spectrum before decompression is 
$f_{\rm sum,k}(p)=f_{\rm inj,k}(p)+ f_{\rm d,k}(p)$,
where $f_{\rm inj,k}$ corresponds to the freshly injected particles 
at the current shock (Equation (\ref{finj})).

After decompression in the downstream region, the re-acceleration spectrum becomes
\begin{equation}
f_{\rm d,k}^{\prime} (p,p_{\rm inj,k})= q_{\rm k} \cdot (p/R_{\rm k})^{-q_{\rm k}} \int_{p_{\rm inj,k}}^{p/R_{\rm k}} t^{q_{\rm k}-1} f_{\rm u,k} (t) dt,
\label{freacc}
\end{equation}
which can be compared with Equation (3) of MP93.
Hereafter, we refer Equation (\ref{freacc}) as the `re-acceleration integration', which in fact
includes both the re-acceleration and decompression steps.
We take the sum of the injection and re-acceleration spectra,
\begin{equation}
f_{\rm sum,k}^{\prime}(p)=f_{\rm inj,k}^{\prime}+ f_{\rm d,k}^{\prime}= f_{\rm u,(k+1)}(p),
\label{fsum}
\end{equation}
as the upstream spectrum at the subsequent shock.
Figure \ref{f1} illustrates this concept.
After decompression, the momentum of some low energy CRs shifts to $p<p_{\rm inj,k}$. They can no longer participate in DSA and join the thermal distribution \citep[e.g.][]{gieseler2000}.

\section{Results}
\label{s3}

We consider the ICM plasma that consists of fully ionized hydrogen atoms and free electrons
with $T_1=5.8\times10^7$~K (5~keV) and $n_H=10^{-4}\cm3$,
so the preshock thermal pressure at the $k$-th shock is $P_{\rm 1,k}=2n_{\rm 1,k}k_B T_1$.
{Note that the normalization of $f(p)$ presented in the figures below scale with $n_H$.

\begin{figure*}[t]
\vskip -0.0cm
\centerline{\includegraphics[width=0.70\textwidth]{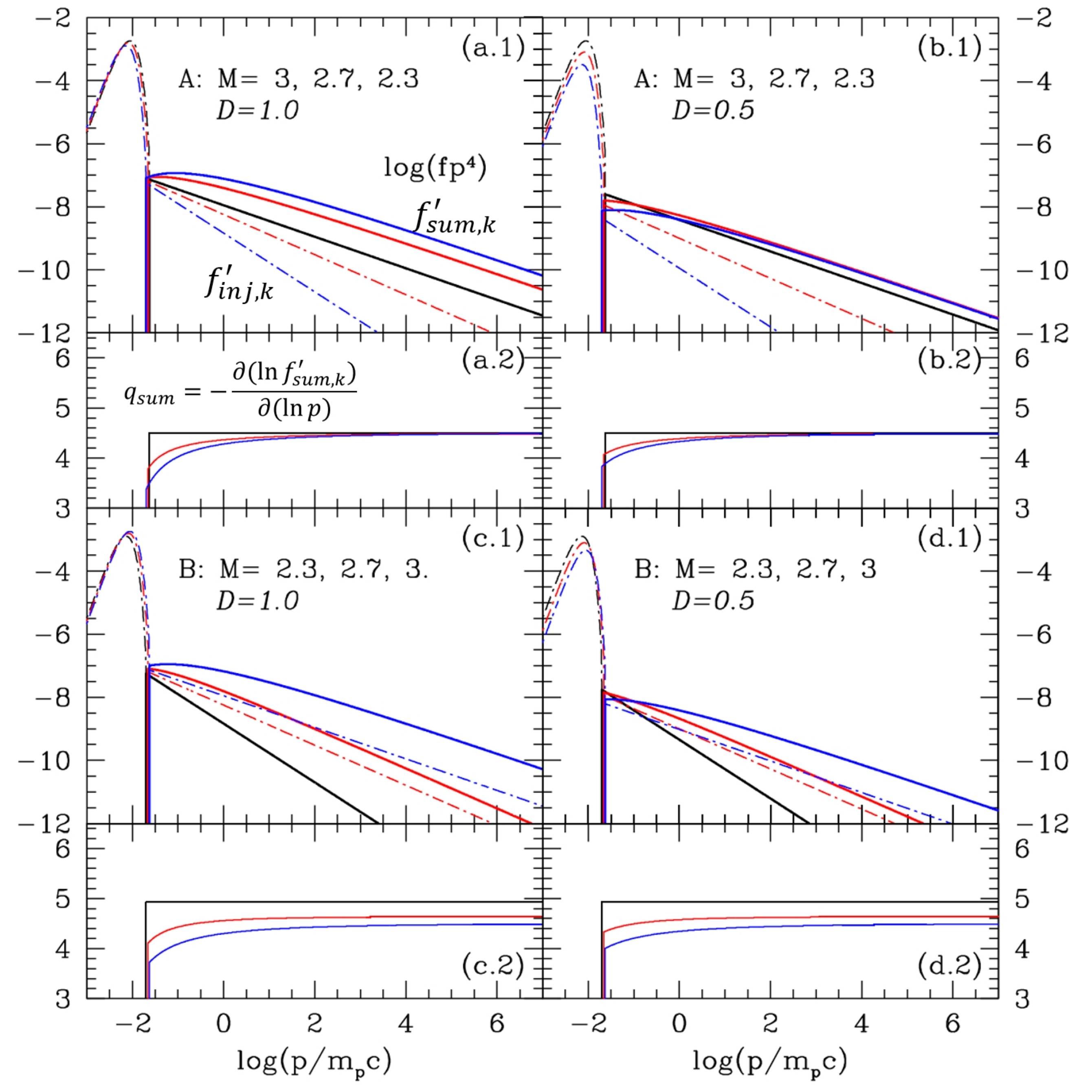}}
\vskip -0cm
\caption{
DSA by a sequence of three shocks with different Mach numbers of $M=$ 3, 2.7, and 2.3.
Two models, A and B, each with $\mathcal{D}=1$ (left panels) and $\mathcal{D}=0.5$ (right panels) are shown.
In Panels (a.1)-(d.1), the dot-dashed lines show $f_{\rm inj,k}^{\prime}$ and the Maxwellian distributions at the $k$-th shock,
while the solid lines show $f_{\rm sum,k}^{\prime}$.
In Panels (a.2)-(d.2), the power-law slope,
$q_{\rm sum}= - \partial ( \ln f_{\rm sum,k}^{\prime})/\partial \ln p$ is shown.
The injection parameter is set to be $Q_{\rm i}=3.8$.
The black, red, and blue lines are used for $k=$1, 2, and 3, respectively.
\label{f3}
}
\end{figure*}

\begin{figure*}[t]
\vskip -0.0cm
\centerline{\includegraphics[width=0.70\textwidth]{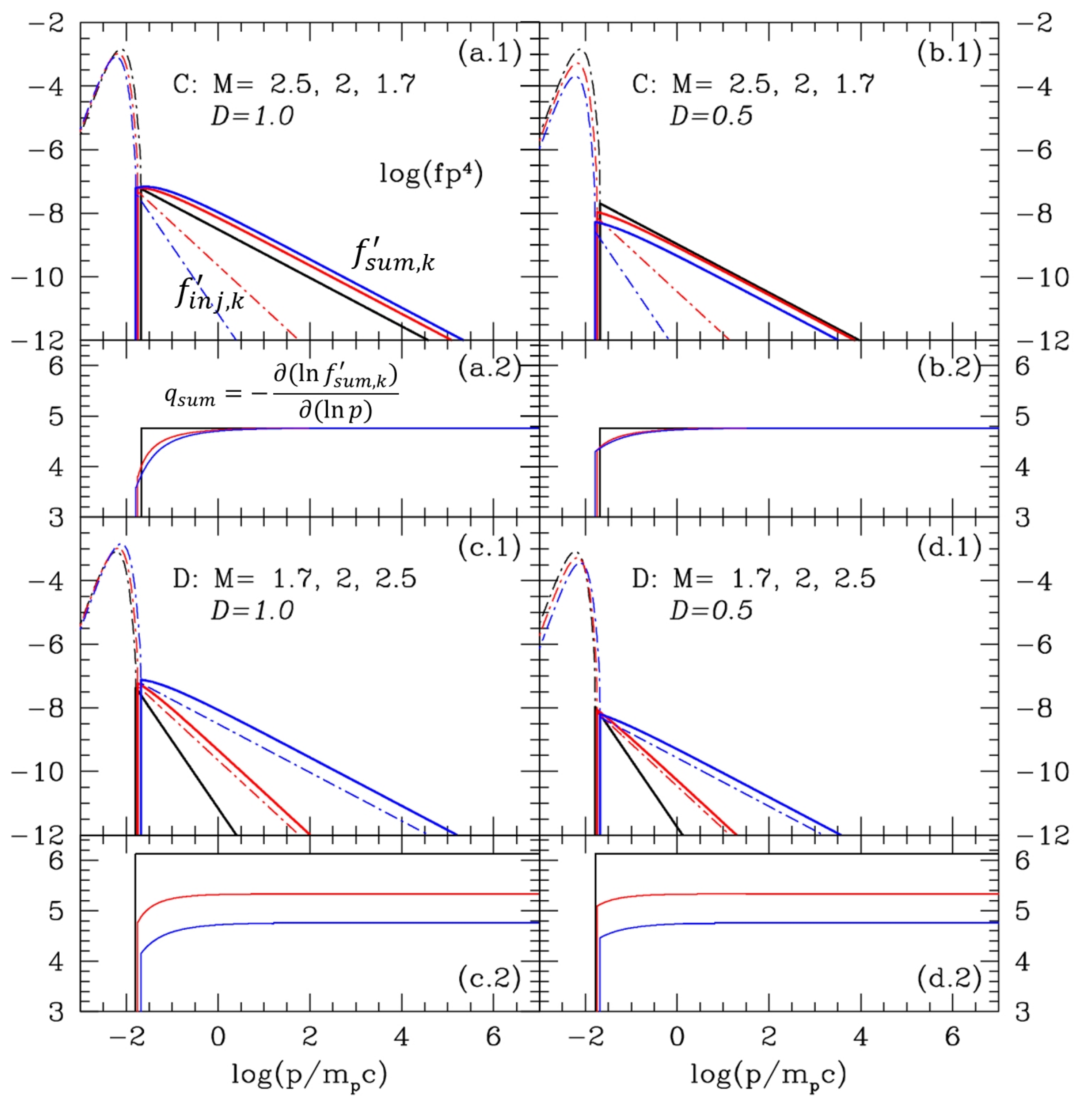}}
\vskip -0.0cm
\caption{
Same as Figure \ref{f3}, except the models C and D with $M=$ 2.5, 2 and 1.7 are shown.
\label{f4}
}
\end{figure*}

\subsection{DSA by Multiple Identical Shocks}
\label{s3.1}

In this section, we consider the special case of multiple {\it identical} shocks 
with the same values of $M$, $q$, $p_{\rm inj}$, and $R$ (without the subscript $k$),
to compare our approach with the analytic treatment of MP93.
Their assumptions differ from ours in the following aspects:
(1) They adopted an injection spectrum with mono-energetic particles, 
$\phi_0(p)=K\delta(p-p_0)$.
In this case, the spectrum of CRs injected and accelerated at the first shock, followed by
the decompression downstream, becomes
\begin{equation}
f_1^{\prime}(p)= {{K~q}\over p_0}\left(p \over {R p_0} \right)^{-q}.
\end{equation}
Note that this is the same as the decompressed injection spectrum 
in Equation (\ref{fpinj}), if we take $K=f_{\rm o,i}p_0/q$ and $p_0=p_{\rm inj}$.
(2) They did not adopt an injection momentum. 
We presume that they set the lower bound of the re-acceleration integration as $p_0R^{k-1}$ (see their Figure 1).
In fact, \citet{gieseler2000} introduced the concept of the injection momentum, $p_0$, 
and made a distinction between thermal and CR distributions. 
So they took $p_0$ as the lower bound of the re-acceleration integration (see their Equation (1)), and considered the CR particles, which are decompressed to $p<p_0$, `re-thermalized'.
(3) MP93 assumed the uniform background medium, so the preshock gas density remains
constant with $\mathcal{D}=1$.

Moreover, MP93 followed the consecutive re-acceleration of injected particles
from the first shock to the $N_s$-th shock (see their Equation (4)).  
Then the contributions from all the shocks were added to obtain the final total spectrum (see their Equation (5)).
By contrast, we add the particles freshly injected/accelerated at each shock 
to the upstream spectrum of the next shock, as illustrated in Figure \ref{f1}.

Here we follow the MP93's approach to derive an approximate analytic expression.
The spectrum of CRs injected/accelerated at the first shock,
after the downstream decompression, is given as
\begin{equation}
f_{\rm 1}^{\prime}(p) = f_{\rm o} \cdot \left(p \over {R p_{\rm inj}} \right) ^{-q}.
\label{finj1}
\end{equation}
Applying the `re-acceleration integration' for the first time at the second shock results in
\begin{equation}
f_{\rm 2 }^{\prime}(p) = f_{\rm o} \cdot q \left(p \over {R^2p_{\rm inj}} \right) ^{-q} 
\ln \left(p\over {Rp_{\rm inj}} \right).
\end{equation}
Then, the far-downstream spectrum behind the $n$-th shock can be found by applying the 
`re-acceleration integration' in Equation (\ref{freacc}) repeatedly for $(n-1)$ times. 
For $p\gg p_{\rm inj}$, the small integration constant at each `re-acceleration integration', 
$\big[ \ln (1/R^{n-1})\big]^{n-1}$, can be ignored, so
the resulting spectrum can be approximated with the following analytic form,
\begin{equation}
f_{\rm n}^{\prime}(p) \approx f_{\rm o} \cdot \frac{q^{n-1}}{(n-1)!} \left(p \over {R^{n} p_{\rm inj}} \right) ^{-q} 
\Big[ \ln \left(p\over {R^{n-1}p_{\rm inj}} \right)\Big]^{n-1}.
\label{fn}
\end{equation}
This equation differs slightly from Equation (4) of MP93, which is an exact expression 
because the integration constant disappears with their lower bound, $p_0R^{k-1}$, of the re-acceleration integration.

Note that $f_{\rm n}^{\prime}(p)$ represents the particles injected at the first shock and
re-accelerated by $(n-1)$ times by subsequent shocks.
Then, the total contribution due to all $N_s$ shocks can be expressed as
\begin{equation}
f_{\rm tot}^{\prime}(p,N_s) \approx \sum_{n=1}^{N_s} f_{\rm n}^{\prime}(p).
\label{ftot}
\end{equation}
In fact, the first term of this summation, $f_{1}^{\prime}(p)$, represents CRs injected/accelerated at the $N_s$-th shock, 
while the last term, $f_{N_s}^{\prime}(p)$, corresponds to CRs injected/accelerated at the first shock and
re-accelerated repeatedly by $(N_s-1)$ consecutive identical shocks.
Note that $f_{\rm tot}^{\prime}(p,N_s)\approx f_{\rm sum,N_s}^{\prime}(p)$, since
the contribution due to the small integration constants can be ignored.

Figure \ref{f2} shows the cases for five identical shocks with $M=3$ (left panels) and $M=20$ (right panels).
Panels (a.1) and (b.1) show the downstream spectrum, $f_{\rm sum,k}^{\prime}(p)$, 
for the cases with $\mathcal{D}=1$ (solid lines) and $\mathcal{D}=0.5$ (dotted lines).
Note that $f_{\rm sum,k}^{\prime}(p)$ is calculated numerically, based on Equation (\ref{fsum}).
Although not shown, the analytic expression, $f_{\rm tot}^{\prime}(p,N_s)$,
approximates closely the numerical estimation, $f_{\rm sum,k}^{\prime}(p)$,
and the difference between the two is small for $p\gg p_{\rm inj}$.
However, the test-particle assumption becomes invalid after several shock passages even with low Mach numbers (see Figure \ref{f6} below).
So the analytic approximation in Equation (\ref{fn}) should be used only for a small number of shocks, e.g., $N_s\lesssim 3-5$.
The cases with $M=20$ are shown for comparisons with the previous studies,
although the test-particle assumption is not valid for such strong shocks.
For example, 
the solid lines in panel (b.2) can be compared with Figure 1 of \citet{gieseler2000}. 

For the parameters relevant for weak ICM shocks, namely, $M= 3$ and $N_s= 3$,
panel (a.2) of Figure \ref{f2} shows that the flattening of the spectrum (blue solid line) 
by multiple shocks could leads to
a substantial deviation from the canonical DSA slope of $q_{\rm DSA}=4.5$.
However, the decompression due to adiabatic expansion ($\mathcal{D}=0.5$, blue dotted line) somewhat 
weakens such flattening effects of multiple shock passages.
On the other hand, if one does not account for the decompression factor of $1/r_{\rm k}$,
the resulting spectrum would be much flatter than shown here.

The results shown in Figure \ref{f2} could be applied as well for low energy CR electrons that 
might not be affected by synchrotron and IC losses.
For example, for radio emitting electrons with the Lorentz factor $\gamma_e\sim 10^4$ ($p/m_pc \sim 5$), the power-law slope, $q_{\rm sum}\approx 4.25$, which is slightly flatter than $q_{\rm DSA}=4.5$ for
a single $M=3$ shock.

\begin{figure*}[t]
\vskip -0.0cm
\centerline{\includegraphics[width=0.7\textwidth]{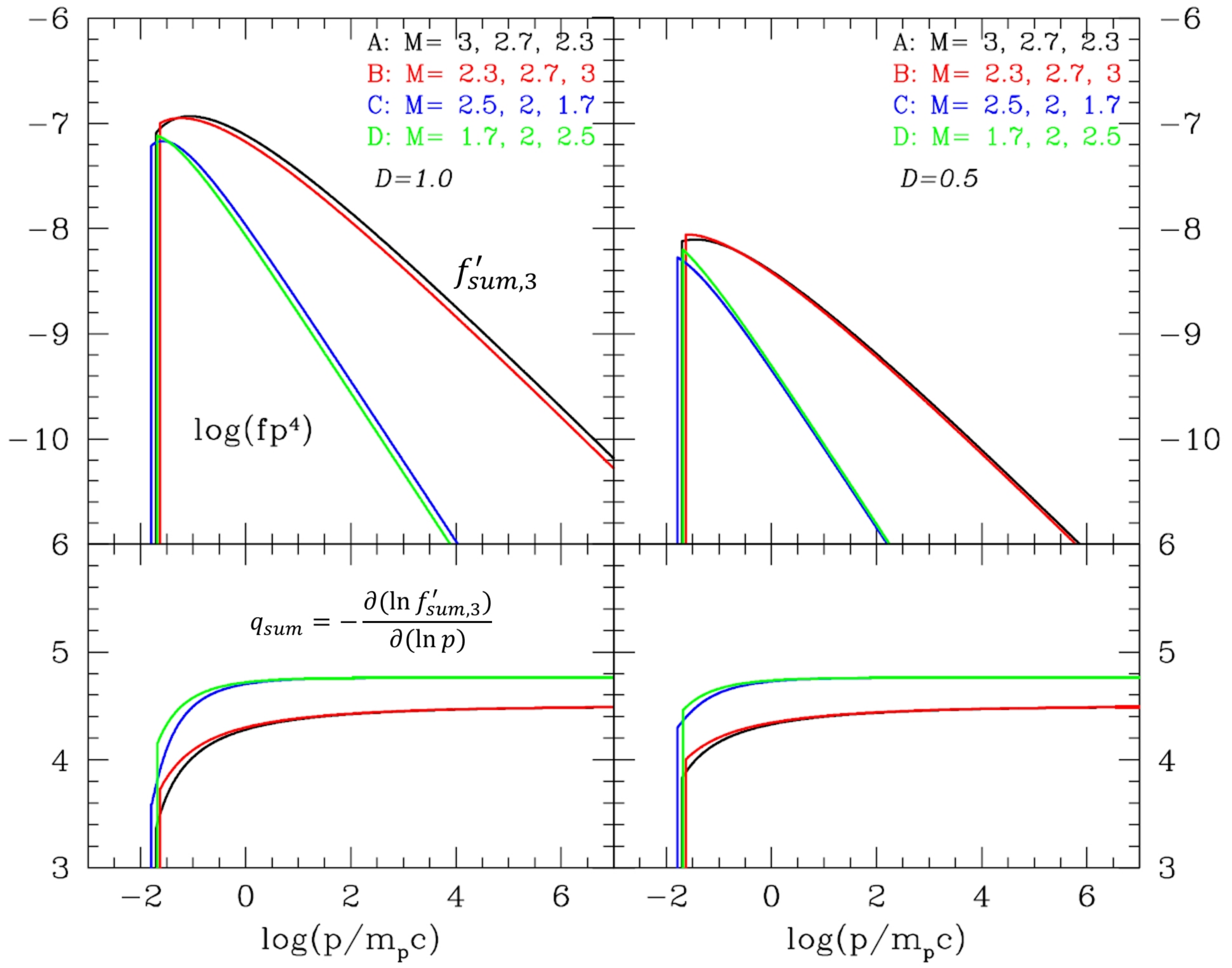}}
\vskip 0.2 cm
\caption{
DSA by a sequence of three shocks with different Mach numbers shown in Figures \ref{f3} and \ref{f4}.
The left (right) panels show the cases with $\mathcal{D}=1$ ($\mathcal{D}=0.5$).
The upper panels show the total spectrum behind the third shock, $f_{\rm sum,3}^{\prime}(p)$,
while the lower panels show $q_{\rm sum}= - \partial ( \ln f_{\rm sum,3}^{\prime})/\partial \ln p$.
The black and red lines compare A and B, while the blue and green lines compare C and D.
\label{f5}
}
\end{figure*}

\begin{figure*}[t]
\vskip -0.0cm
\centerline{\includegraphics[width=0.8\textwidth]{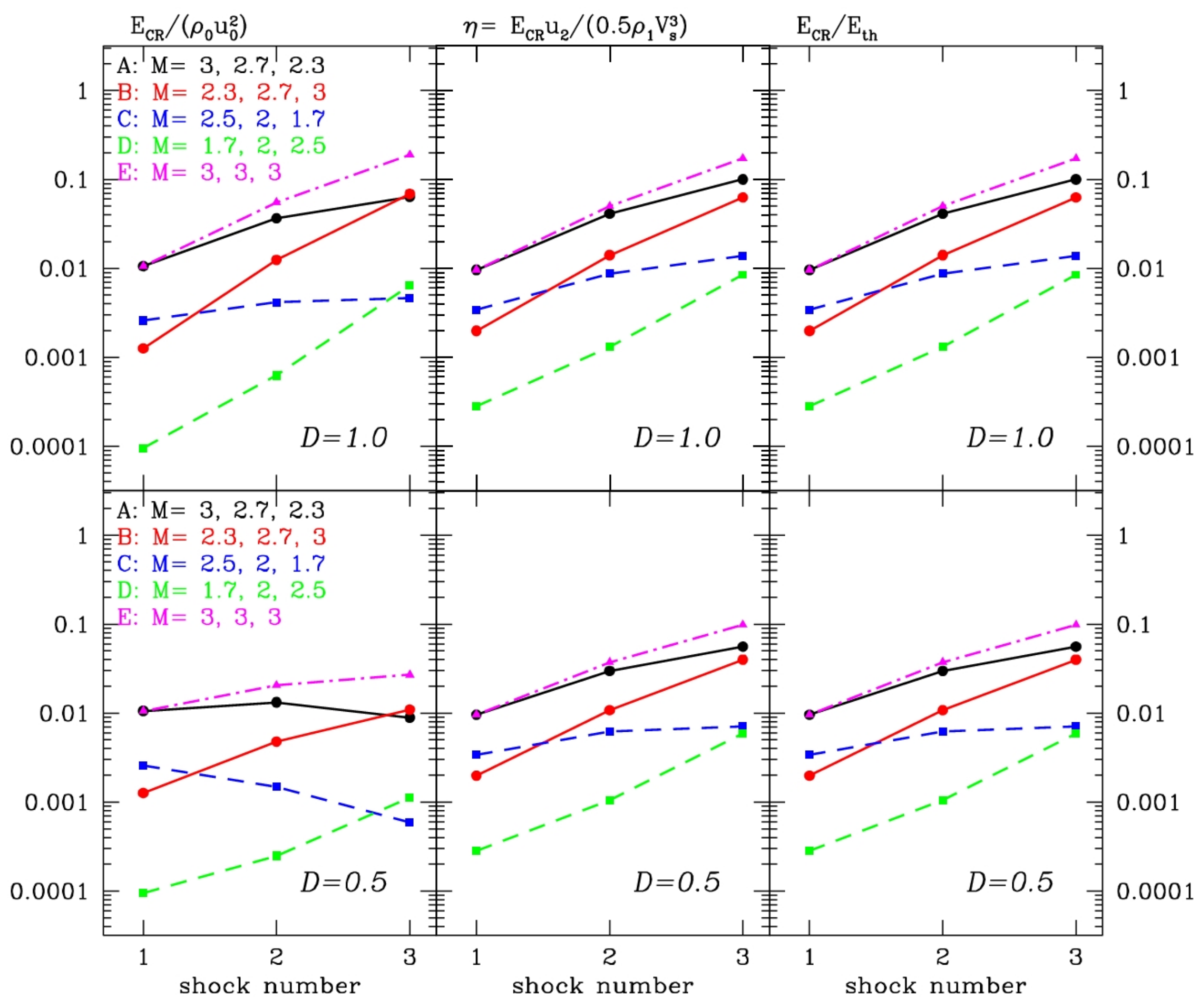}}
\vskip 0.3 cm
\caption{
CR pressure, $E_{\rm CR,k}/\rho_0 u_0^2$,
the DSA efficiency, $\eta$, and the ratio, $E_{\rm CR,k}/E_{\rm th,k}$, due to a sequence of three shocks,
where $E_{\rm CR,k}$ and $E_{\rm th,k}$ are the CR and thermal energy densities in the immediate postshock region at the $k$-th shock.
The constant normalization factor, $\rho_0 u_0^2$, corresponds to the shock kinetic energy density for the preshock gas density, $\rho_0$, and the shock speed, $u_0$, with $M=3$. 
Five cases, A-E, with different combinations of $M$'s are shown.
The upper (lower) panels show the cases with $\mathcal{D}=1$ ($\mathcal{D}=0.5$).
\label{f6}
}
\end{figure*}

\subsection{DSA by Multiple Shocks with Different $M_s$}
\label{s3.2}

We now explore a more realistic scenario, in which weak ICM shocks with different Mach numbers 
of $M\lesssim 3$.
Figure \ref{f3} shows the two cases, A ($M=$ 3, 2.7, 2.3 in upper panels) and B ($M=$ 2.3, 2.7, 3 in lower panels) 
with $\mathcal{D}=1$ (left panels) and $\mathcal{D}=0.5$ (right panels).
Figure \ref{f4} compares the two cases,  C ($M=$ 2.5, 2, 1.7 in upper panels) and D 
($M=$ 1.7, 2, 2.5 in lower panels).  
In panels (a.1)-(d.1), the solid lines show the total spectrum, $f_{\rm sum,k}^{\prime}$ in the far-downstream region of the $k$-th shock (where the index $k$ runs from 1 to 3),
while the dashed lines show the injection spectrum, $f_{\rm inj,k}^{\prime}$,
injected/accelerated at the $k$-th shock and decompressed far-downstream.
Note that for the first shock, $f_{\rm sum,1}^{\prime}=f_{\rm inj,1}^{\prime}$, so the black dashed line
overlap with the black solid line.

The comparison of A and B (or C and D) illustrates that the final spectrum (blue solid lines)
is determined by the accumulated effects of repeated re-acceleration by multiple shocks,
but almost independent of the ordering of Mach numbers.
The black and red lines in Figure \ref{f5} compare the final spectrum at the third shock, $f_{\rm sum,3}^{\prime}$ for A and B, while the blue and green lines compare C and D. 
The amplitudes of $f_{\rm sum,3}^{\prime}$ differ slightly due to the small
difference in $p_{\rm inj,k}$ and $f_{\rm o,k}$ at each shock. 
Otherwise, the slopes, $q_{\rm sum}$, are almost the same for $p/m_pc > 1$ in the two cases,
A and B (or C and D).

In the DSA theory in the test-particle regime, the mean momentum gain of a particle at each
shock crossing is $\langle \Delta p \rangle = (2/3) p (u_1 - u_2)/v$,
where $u_1$ and $u_2$ are the preshock and postshock flow speeds, respectively,
in the shock rest frame, and $v$ is the particle speed.
The probability of the particle returning from downstream is $P_{\rm ret} = (1-4u_2/v)$.
These two factors, $\langle \Delta p \rangle$, and $P_{\rm ret}$, determine
the DSA power-spectrum, $f(p)\propto p^{-q}$ \citep{drury1983}.
The momentum gain of a particle through repeated shock crossings in a sequence of multiple 
shocks is cumulative, so it should not depend on the ordering of the shock Mach numbers.
On the other hand, the amount of injected particles at each shock depends on $M_{\rm k}$ in our injection
scheme, since $p_{\rm inj,k}$ and $f_{\rm o,k}$ depend on the postshock temperature $T_{\rm 2,k}$
and the density of the incoming proton density, $n_{\rm 1,(k+1)}=D n_{\rm 1,k}$, may decrease at subsequent shocks.
For example, case A with $M_1=3$ injects more particles at the first shock than case B with
$M_1=2.3$, and hence the resulting spectrum has a slightly higher amplitude.
As illustrated in Figure \ref{f5}, the final CR spectrum produced from DSA by multiple shocks is 
almost independent of the ordering of $M$'s, other than small differences in the amplitude
due to different injection rates at each shock.

Figure \ref{f5} also shows that, in the cases with $\mathcal{D}=0.5$ with additional decompression, 
the amplitudes of $f_{\rm sum,3}^{\prime}$ are smaller  
and the slopes, $q_{\rm sum}$, are flattened slightly less than the cases with $\mathcal{D}=1$.
Thus, the additional decompression parameterized with $\mathcal{D}$ weakens the flattening effects
of multiple shocks.

\subsection{DSA Efficiency}
\label{s3.3}

In this section, we examine how multiple passages of shocks affect the DSA efficiency.
First, we have calculated the CR energy density in the {\it immediate postshock} region behind the $k$-th shock,
\begin{equation}
E_{\rm CR,k} = 4 \pi c \int_{p_{\rm inj,k}} (\sqrt{p^2+ (m_pc)^2}-m_pc) f_{\rm sum,k}(p) p^2 dp,
\label{ECR}
\end{equation}
where $f_{\rm sum,k}(p)=f_{\rm inj,k} + f_{\rm d,k}$ is the postshock spectrum {\it before decompression}.
The DSA efficiency is commonly defined in terms of the shock kinetic energy flux  \citep{ryu2003}, 
\begin{equation}
\eta\equiv {{E_{\rm CR,k} u_{\rm 2,k}}\over {0.5\rho_{1,k} V_{\rm s,k}^3}},
\end{equation}
where $u_{\rm 2,k}$ is the postshock flow speed and the denominator is the incident kinetic energy flux of the $k$-th shock.
Another measure for the DSA efficiency can be given in terms of the ratio, $E_{\rm CR,k}/E_{\rm th,k}$,
where $E_{\rm th,k}$ is the postshock thermal energy density behind the $k$-th shock.

Figure \ref{f6} shows $E_{\rm CR,k}/\rho_0 u_0^2$ (left column), $\eta$ (middle column) and the ratio of $E_{\rm CR,k}/E_{\rm th,k}$ (right column) at each shock for the five cases, A-E, 
with $\mathcal{D}=1$ (upper panels) and $\mathcal{D}=0.5$ (lower panels).
Note that the normalization factor, $\rho_0 u_0^2$, is a constant, which corresponds to the
shock kinetic energy density for the preshock gas, $\rho_0$, and the shock speed, $u_0$, with $M=3$. 
The three quantities increase with the successive passages of shocks, as expected.
The comparison of A (black) and B (red) or the comparison of C (blue) and D (green) demonstrate 
that the final efficiencies at the third shock do depend slightly on the ordering of the three values of $M_{\rm k}$'s.
This may seem inconsistent with Figure \ref{f5}.
However, we note that $E_{\rm CR}$ is estimated from the immediate postshock spectrum, $f_{\rm sum,k}$, before decompression,
while Figure \ref{f5} shows the far-downstream spectrum, $f_{\rm sum,3}^{\prime}$ decompressed in the downstream region. 
In addition, both the denominator in $\eta$, $0.5\rho_{\rm 1,k} V_{\rm s,k}^3$,
and the denominator in the ratio, $E_{\rm th,k}$, do depend on $M_{\rm k}$ of the $k$-th shock. 
So, in the cases with higher $M_3$ (B and D), $\eta$ and $E_{\rm CR,3}/E_{\rm th,3}$ are smaller
at the third shock, compared with the cases with lower $M_3$ (A and C).

Obviously, case E (magenta) with three identical shocks with $M=3$ and $\mathcal{D}=1$ produces the highest amount of CRs.
So $\eta$ increases as $7.1\times 10^{-3}$, $3.7\times 10^{-2}$, and $0.13$ for $k=$ 1, 2, and 3,
respectively,
while $E_{\rm CR,k}/E_{\rm th,k}$ increases as $9.6\times 10^{-3}$, $5.0\times 10^{-2}$, and $0.17$.
The other four cases with different $M$'s, A-D, $\eta \lesssim 0.1$ after three shock passages, 
so the test-particle assumption remains valid.

The cases with $\mathcal{D}=0.5$ with additional decompression (lower panels) produce 
smaller $E_{\rm CR}$ than the cases with $\mathcal{D}=1$ (upper panels).
Note that the ratio, $E_{\rm CR,k}/\rho_0 u_0^2$, could decrease at subsequent shocks 
because of the successively lower preshock density, $n_{\rm 1,k}$.

\section{Summary}
\label{s4}

We have revisited the problem of DSA by multiple shocks, considering weak ICM shocks in the test-particle regime.
Suprathermal particles with $p\ge p_{\rm inj}\sim 3.8 p_{\rm th}$ are assumed to be injected to the DSA process at each shock,
while incident upstream CRs are re-accelerated by subsequent shocks \citep{ryu2019,kang2020}.
Based on Liouville's theorem, the accelerated CRs are assumed to undergo adiabatic decompression by a factor of
$R=(D/r)^{1/3}$, which takes into account the expansion of the background medium
and the decompression of the postshock flow behind each shock (see MP93).
The decompression causes the particle momentum, $p$, to decrease to $p^{\prime}=Rp$.

We have considered the several examples with three shocks with the sonic Mach numbers, $M = 1.7 - 3$.
The main findings can be summarized as follows:

\begin{enumerate}

\item Multiple passages of {\it identical} shocks with the same Mach number enhance 
the amplitude of the momentum distribution, $f_{\rm sum,k}^{\prime}(p)$,
and flatten the slope, $q_{\rm sum}= - \partial ( \ln f_{\rm sum,k}^\prime)/\partial \ln p$
(see Figure \ref{f2}).
This is in a good agreement with the previous studies \citep[e.g.,][]{schneider1993,melrose1993,gieseler2000}.
Moreover, the DSA efficiency, $\eta$ and the ratio $E_{\rm CR,k}/E_{\rm th,k}$ increase during each shock passage (see Figure \ref{f6}).

\item For the cases with multiple shocks with different Mach numbers,
the differences in the injection momentum, $p_{\rm inj,k}$, and the ensuing normalization factor,
$f_{\rm o,k}$, at each shock lead to small differences in the final CR spectrum, $f_{\rm sum,3}^{\prime}(p)$, depending on the ordering of $M$'s (see Figures \ref{f3} and \ref{f4}).
The final spectrum of CRs at the last shock turns out to be 
almost independent of the ordering of $M$'s (see Figure \ref{f5}).
Likewise, the power-law slope, $q_{\rm sum}(p)$, remains almost unchanged as long as $p/m_pc \gg 1$, regardless of the order of $M$'s.

\item 
The power-law slope of final CR spectrum is governed by the accumulated effects of repeated re-acceleration by past shocks, meaning it can be different from the DSA power-law slope,
$q_{\rm DSA}$, of the current shock.

\item The decompression in the downstream region of each shock due to the expansion of the background
medium weakens the amplification and flattening of the resulting CR spectrum.

\end{enumerate}

Although we have considered DSA of CR protons in this study, the same approach can be applied to the 
CR electron spectrum injected and re-accelerated at multiple shocks, 
if the energy losses due to synchrotron emission and iC scattering are taken into account.
For example, if the timescale of shock passage is comparable to or greater than the synchroton cooling time,
$t_{\rm sync}\approx 10^8 \yrs (B_{\rm eff}/5\muG)^{-2}(\gamma_e/10^4)^{-1}$,
the flattening and amplifying effects of multiple shock passages would be reduced substantially. 
The effective magnetic field strength, $B_{\rm eff}$, takes into account both
synchrotron and iC cooling, and $\gamma_e$ is the Lorentz factor of radio-emitting electrons.
We leave such calculations including treatments of the energy losses to a upcoming paper.

We recognize the possibility that the re-acceleration by multiple shocks may
explain the discrepancies, $M_{\rm x} \lesssim M_{\rm radio}$, found in some radio relics \citep{akamatsu13,hong2015,vanweeren2019}, as discussed in the Introduction.
In other words, $M_{\rm x}$, inferred from X-ray observations (e.g., temperature jump) may represent the Mach number 
of the most recent shock, while $M_{\rm radio}$, inferred from radio observations (e.g., ratio spectral index), may 
reflect the accumulated effects of all past shocks.

\acknowledgments
The author thanks the anonymous referee for constructive comments.
This work was supported by a 2-Year Research Grant of Pusan National University.


\end{document}